\documentclass[useAMS,usenatbib]{mn2e}
\usepackage{graphicx}
\usepackage{txfonts}
\usepackage{natbib}
\def\ms{{\rm M_{\odot}}}
\def\be{\begin{equation}}
\def\ee{\end{equation}}
\def\bea{\begin{eqnarray}}
\def\eea{\end{eqnarray}}

\def\ms{{\rm M_{\odot}}}

\def\rsix{R_{6}}

\def\po{\ifmmode P_{o} \else $P_{o}$ \fi}

\def\la{ \langle}
\def\ra{ \rangle}

\def\ga{\gamma}

\def\la{\lambda}

\def\mdot{\ifmmode \dot M \else $\dot M$\fi}    % accretion rate
\def\mxd{\ifmmode \dot {M}_{x} \else $\dot {M}_{x}$\fi}
\def\med{\ifmmode \dot {M}_{Edd} \else $\dot {M}_{Edd}$\fi}
\def\bff{\ifmmode B_{{\rm f}} \else $B_{{\rm f}}$\fi}

\def\apj{\ifmmode ApJ \else ApJ \fi}    % lower
\def\apjl{\ifmmode  ApJ \else ApJ \fi}    %
\def\aap{\ifmmode A\&A \else A\&A\fi}    %
\def\mnras{\ifmmode MNRAS \else MNRAS \fi}    %
\def\nat{\ifmmode Nature \else Nature \fi}
\def\prl{\ifmmode Phys. Rev. Lett. \else Phys. Rev. Lett.\fi}
\def\prd{\ifmmode Phys. Rev. D. \else Phys. Rev. D.\fi}

\def\ms{\ifmmode M_{\odot} \else $M_{\odot}$\fi}    % lower
\def\na{\ifmmode \nu_{A} \else $\nu_{A}$\fi}    % Alfven frequency
\def\nk{\ifmmode \nu_{K} \else $\nu_{K}$\fi}    % Keplerian frequency
\def\ns{\ifmmode \nu_{{\rm s}} \else $\nu_{{\rm s}}$\fi}
\def\no{\ifmmode \nu_{1} \else $\nu_{1}$\fi}    % lower
\def\nt{\ifmmode \nu_{2} \else $\nu_{2}$\fi}    % upper
\def\ntk{\ifmmode \nu_{2k} \else $\nu_{2k}$\fi}    % upper
\def\dnmax{\ifmmode \Delta \nu_{max} \else $\Delta \nu_{2max}$\fi}
\def\ntmax{\ifmmode \nu_{2max} \else $\nu_{2max}$\fi}    % upper
\def\nomax{\ifmmode \nu_{1max} \else $\nu_{1max}$\fi}    % upper
\def\nh{\ifmmode \nu_{\rm HBO} \else $\nu_{\rm HBO}$\fi}    % HBO
\def\nqpo{\ifmmode \nu_{QPO} \else $\nu_{QPO}$\fi}    % HBO
\def\nz{\ifmmode \nu_{o} \else $\nu_{o}$\fi}    % HBO
\def\nht{\ifmmode \nu_{H2} \else $\nu_{H2}$\fi}    % HBO
\def\ns{\ifmmode \nu_{s} \else $\nu_{s}$\fi}    % stellar
\def\nb{\ifmmode \nu_{{\rm burst}} \else $\nu_{{\rm burst}}$\fi}
\def\nkm{\ifmmode \nu_{km} \else $\nu_{km}$\fi}    % stellar
\def\ka{\ifmmode \kappa \else \kappa\fi}    % stellar
\def\dn{\ifmmode \Delta\nu \else \Delta\nu\fi}    % stellar

\def\rs{\ifmmode {R_{s}} \else $R_{s}$\fi}    % stellar
\def\ra{\ifmmode R_{A} \else $R_{A}$\fi}    % Alfven radius
\def\rso{\ifmmode R_{S1} \else $R_{S1}$\fi}    % sonic point radius
\def\rst{\ifmmode R_{S2} \else $R_{S2}$\fi}    % sonic point radius
\def\rmm{\ifmmode R_{M} \else $R_{M}$\fi}    % stellar
\def\rco{\ifmmode R_{co} \else $R_{co}$\fi}    % stellar
\def\ris{\ifmmode {R}_{{\rm ISCO}} \else $ {\rm R}_{{\rm ISCO}} $\fi}
\def\rsix{\ifmmode {R_{6}} \else $R_{6}$\fi}
\def\rinfty{\ifmmode {R_{\infty}} \else $R_{\infty}$\fi}
\def\rinfsix{\ifmmode {R_{\infty6}} \else $R_{\infty6}$\fi}

\def\rxj{\ifmmode {RX J1856.5-3754} \else RX J1856.5-3754\fi}
\def\1739{\ifmmode {XTE  J1739-285} \else XTE  J1739-285\fi}
\def\exo{\ifmmode {EXO 0748-676} \else EXO 0748-676\fi}
\def\m18{{\rm{\dot M_{18}}}}

\title[Magnetic field evolution of ULX NuSTAR J095551+6940.8]
{%The Magnetic Field and Spin Period Evolution  of
The magnetic field evolution of ULX NuSTAR J095551+6940.8 in M82---a legacy of accreting magnetar}

\author[Y. Y. Pan et al.]{Y.Y. Pan$^{1,2}$ \thanks{E-mail:
panyy@ihep.ac.cn, zhangcm@bao.ac.cn}, L.M. Song$^{1}$, C. M. Zhang$^{2,3}$, H. Tong$^{4}$  \\
$^{1}$Key Laboratory of Particle Astrophysics, Institute of High Energy Physics, Chinese Academy of Sciences, Beijing 100049, China\\
$^{2}$National Astronomical
Observatories, Chinese Academy of Sciences, Beijing 100012,
China\\
  $^3$ Key Laboratory of Radio astronomy, Chinese Academy of Sciences, Beijing 100012, China\\
$^{4}$Xinjiang  Astronomical Observatory, Chinese Academy of
Sciences, Urumqi 830011, China}
\begin{document}
\date{Accepted Date; Received Date}

\pagerange{\pageref{firstpage}--\pageref{lastpage}} \pubyear{}

\maketitle

\label{firstpage}

\begin{abstract}

Ultra-luminous X-ray sources are usually believed to be black holes with mass about $10^{2-3}M_{\odot}$. However, the recent discovery of NuSTAR J095551+6940.8 in M82 by Bachetti et al. shows that it holds the spin period $P=1.37\rm\,s$ and period derivative $\dot{P}\approx-2\times10^{-10}\rm\,s\,s^{-1}$, which provides a strong evidence that some ultra-luminous X-ray sources could be neutron stars.
We obtain that the source may be an evolved magnetar according to our simulation by employing the model of accretion induced the polar magnetic field decay and standard spin-up torque of an accreting neutron star. The results show that NuSTAR J095551+6940.8 is still in the spin-up process, and the polar magnetic field decays to about $4.5\times10^{12}\rm\,G$ after accreting $\sim 10^{-2.5}$\ms, while the strong magnetic field  exists in the out-polar region, which could be responsible for the observed low field magnetar. The ultra luminosity of the source can be explained by the beaming effect and two kinds of accretion--radial random accretion and disk accretion. Since the birth rate of magnetars is about ten percent of the normal neutron stars, we guess that several ultra-luminous X-ray sources should share the similar properties to that of NuSTAR J095551+6940.8.

\end{abstract}


\section{Introduction}

Ultra-luminous X-ray sources (ULXs), also named super-Eddington sources or super-luminous sources, were discovered by Einstein Observatory in 1980's (Fabbiano 1989). They are usually thought as the point sources with luminosity about $10^{39}\rm\,erg\,s^{-1}$ or above in the accreting systems. Assuming that the isotropic emission is with Eddington limit, ULXs may be black holes whose mass go beyond those of the stellar black holes (Makeshima et al. 2000; Feng and Soria 2011; Pasham et al. 2014; Weng et al. 2014).
Recently, an ULX NuSTAR J095551+6940.8 in the nearby galaxy M82 was discovered by Bachetti et al. (2014), which may be the counterpart of NuSTAR J095551+6940.8 (Feng \& Kaaret 2007; Bachetti et al. 2014). The spin period ($P=1.37\rm\,s$) and period derivative ($\dot{P}\approx-2\times10^{-10}\rm\,s\,s^{-1}$) prove that it is an accreting neutron star (NS) in high mass binary system (HMXB) (Bachetti et al. 2014), which sheds a doubt on that all ULXs are accreting black holes.

Studying on NuSTAR J095551+6940.8 would help us understand various distinctive characteristics of ULXs, and how an accreting neutron star can be explained as an ULX. To this point, many researchers have paid much attentions.
Shao and Li (2015) modeled the formation history of NuSTAR J095551+6940.8. They suggested that
NS in X-ray binary system might significantly contribute to ULX population. High-mass and intermediate-mass X-ray binary pulsars dominated the ULX population in M82 and Milky way like galaxies, respectively.
Klu$\rm \acute{z}$niak and Lasota (2015) pointed out that NuSTAR J095551+6940.8 would become a millisecond pulsar within $10^5\rm\,yr$, which provided a new way of millisecond pulsar formation in HMXBs.
Dall'Osso et al. (2015) proposed that the favorite magnetic field of NuSTAR J095551+6940.8 was $\sim10^{13}\rm\,G$ based on the physical considerations and the observed properties of the source.
Mushtukov et al. (2015) pointed out that the surface magnetic field strength was $\sim10^{14}\rm\,G$ for NuSTAR J095551+6940.8 with the observed luminosity $\sim10^{40}\rm\,erg\,s^{-1}$, but could not be too large due to the propeller effect.
Eksi et al. (2015) showed that its dipole magnetic field was as strong as $6.7\times10^{13}\rm\,G$ according to the torque equilibrium condition.
As expected, it could hold even stronger multi-pole field that suppresses the scattering cross section and increases the Eddington luminosity (Eksi \& Alpar 2003, Canuto et al. 1971).
Moreover, Tong (2015a, 2015b) proposed that NuSTAR J095551+6940.8 might be a magnetar in an accreting system, which favored the idea that magnetars were descendants of high mass X-ray binaries (Bisnovatyi-Kogan \& Ikhsanov 2015).

Magnetar is always an isolated NS with the super-strong magnetic field about $10^{14-15}\rm\,G$,
which is always thought to be the candidate of soft gamma repeater or anomalous X-ray pulsar (Duncan \& Thompson 1992; Kouveliotou et al. 1998; Mereghetti 2008; Bisnovatyi-Kogan \& Ikhsanov 2014; Olausen \& Kaspi 2014). Until now, there is no evidence that indicates magnetar exists in binary system, though it could not be excluded from the theoretical ground. It has been proposed that the accreting magnetar in the close binary system could be used for interpreting the slow rotating NS in 4U 2206+54 (Ikhsanov \& Beskrovnaya 2010). However, magnetars in binary systems should be the rare case, since the birth rate of magnetar is only ten percent of the normal NS (Lorimer 2008).
Therefore, it can be imagined that a successful explanation of the status of NuSTAR J095551+6940.8 can help us understand NS as an ULX that exists in binary system.

By means of the accretion induced NS magnetic field evolution model (Zhang \& Kojima 2006) and NS spin-up formula by Ghosh and Lamb (1979), we calculated the evolution of the magnetic field and spin period of a magnetar with the appropriate initial condition to make the evolved spin period fit the status of NuSTAR J095551+6940.8. With the result, we compare its evolutionary track of the magnetic field and spin period with those of the observed pulsar in binary systems. The paper is organized as below: the model is described in Section 2; calculations and B-P diagram are presented in Section 3; and the discussion and summary are included in section 4.

\section{The Model}

The model of accretion induced magnetic field decay of a
NS was proposed by Zhang and Kojima (2006). It is assumed that the
accretion material is channeled by the strong magnetic field to
the polar cap, where the field lines are expelled towards the
out-polar regime and enter into the magnetic equator region. As a
result, the strength of the polar field decays and the piled-up
stronger out-polar field lines at equator region are mostly
squeezed into the inner crust of the NS. The analytical form of
the polar field decay of a NS is obtained,
\be B = \frac{\bff}{\{1 - [C/\exp(y)-1]^2\}^{\frac{7}{4}}}\,,
\label{zk} \ee
%%%%%
where $y=2\Delta M/7M_{\rm cr}$ is a ratio parameter of the accretion mass $\Delta M$ and the crust mass $M_{\rm cr}\sim 0.2\ms$ of a NS. $C=1+(1-X^2_0)^{\frac{1}{2}}\sim 2$ is with $X^2_0=(B_{\rm f}/{B_0})^{4/7}$, where $B_0$ is the initial field at the beginning of the accretion, and $B_f$ is the bottom magnetic field while the magnetosphere is compressed onto the surface of NS,
%%%%
\be
B_f=4.5\times 10^8{\rm\,(G)}\cdot \dot M_{18}^{\frac{1}{2}}m^{\frac{1}{4}}R_6^{-\frac{5}{4}}(\phi/0.5) ^{-\frac{7}{4}},
\label{bf}
\ee
%%%
where $\dot M_{18}$, m and $R_6$ are the NS accretion rate in units of $10^{18}\rm\,g\,s^{-1}$, the mass in solar mass and radius in units of $10^6\rm\,cm$, respectively. In detail, with the condition of the magnetosphere radius $R_{\rm M}$ matching the NS radius R, i.e., $R_{\rm M}=R$, the bottom field $B_{\rm f}$ is achieved. Generally, $R_{\rm M}$ is related to the Alfv$\acute{e}$n radius $R_{\rm A}$
\be
R_{\rm M}=\phi R_{\rm A}\;,\; \;
R_{\rm A}=3.2\times10^8{\rm\,(cm)}\cdot \dot M_{17}^{-\frac{2}{7}}\mu_{30}^{\frac{4}{7}}m^{-\frac{1}{7}},
\label{ra}
\label{rm}
\ee
where $\mu_{30}$ is the magnetic moment in units of $10^{30}\rm\,g\,cm^{3}$ and the preferred parameter $\phi\sim0.5$ is usually considered (Ghosh \& Lamb 1979; Shapiro \& Teukolsky 1983; Frank et al. 2002).

The evolution of the spin period of a NS in the accretion phase described by Ghosh \& Lamb (1979) is
\bea
-\dot{P}=5.8\!\times\!10^{-5}{\rm\,(s\,yr^{-1})\,}\!\cdot\! [m^{-\frac{3}{7}}R_6^{\frac{12}{7}}I_{45}^{-1}]B_{12}^{\frac{2}{7}}
(PL_{37}^{\frac{3}{7}})^2n(\omega_{\rm s}),
\label{pdot1}
\eea
where $I_{45}$ is the moment of inertia in units of $10^{45}\rm\,g\,cm^2$ and $L_{37}$ is luminosity of the accreting NS in units of $10^{37}\rm\,erg\,s^{-1}$; $B_{12}$ is the polar magnetic field in units of $10^{12}\rm\,G$; The dimensionless torque $n(\omega_{\rm s})$ is
\be
n(\omega_{\rm s}) = 1.4\times \left(\frac{1-\omega_{\rm s}/\omega_{c}}{1-\omega_{\rm s}} \right),
\label{ns}
\ee
which is associated with the fastness parameter $\omega_{\rm s}\equiv\Omega_{\rm s}/\Omega_{\rm k}$, where $\Omega_{\rm s} = 2\pi/P$ is the spin frequency of the NS. $\Omega_k$ is the Keplerian frequency at the radius $R_{\rm M}$, $\omega_{\rm c}$ is the critical value in the range of $0.2-0.9$ (Ghosh \&  Lamb 1992).

During the NS spin-up process, the spin frequency will be constrained by the Keplerian orbital frequency of the accreting material at the magnetosphere radius. Therefore, the minimum spin period corresponding to the field strength can be described by the equation of the spin-up line (or equilibrium period line) with condition of the spin frequency being the same as the Keplerian frequency at the magnetosphere radius (Bhattacharya and van den Heuvel 1991):
\bea
P_{\rm eq}=0.89\,{(\rm s)}\,\cdot B^{6/7}_{12}m^{-5/7}\dot{M}_{18}^{-3/7}R_6^{16/7}.
\label{spin}
\eea
The corresponding line of Eq.(\ref{spin}) has been plotted as the solid line in Fig.\ref{md1850} with the Eddington rate and standard NS parameters, e.g. $m=1.4\ms$ and $R=10\rm\,km$.

\section{The Magnetic Field of N{\MakeLowercase u}STAR J095551+6940.8}

As an accreting NS, NuSTAR J095551+6940.8 is a unique source since its luminosity and period variation ($L\sim10^{40}\rm\,erg\,s^{-1}$ and $\dot{P}\approx-2\times10^{-10}\rm\,s\,s^{-1}$) are higher than that of NS in HMXBs ($L\sim10^{36-38}\rm\,erg\,s^{-1}$ and $\dot{P}\sim10^{-12}\rm\,s\,s^{-1}$) (Ghosh \& Lamb 1979; Shapiro \& Teukolsky 1983; Bildsten et al. 1997; Frank et al. 2002; Liu et al. 2007).
So its evolution history may be different from that of the usual X-ray pulsars in HMXBs.
The low field magnetar SGR 0418+5729 had been discovered with the dipole magnetic field less than $7.5\times10^{12}\rm\,G$ and the strong field of $B\sim10^{14}$-$10^{15}\rm\,G$ which might be internal (Rea et al. 2010, 2012; Turolla et al. 2011).
Therefore, we simulate the magnetic field and spin period evolution of a magnetar based on the model of accretion induced polar magnetic field decay of a NS (Zhang \& Kojima 2006).
During the simulation, we set the initial condition (accretion rate, initial magnetic field and spin period) and try to find a magnetar with the evolved values of spin period and period derivative to be the same as those of NuSTAR J095551+6940.8 that is spinning up now.

\begin{figure}
\includegraphics[width=8cm]{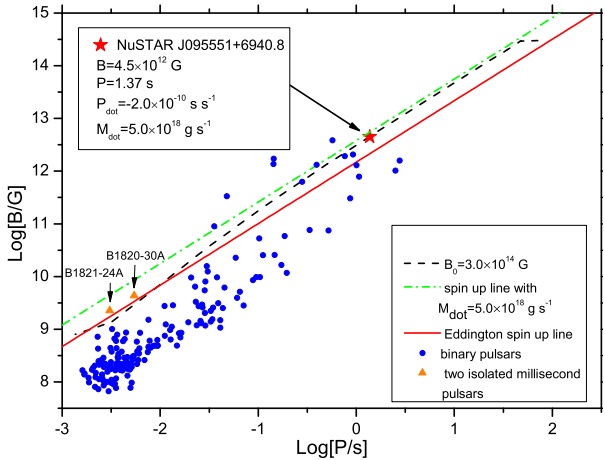}
\caption{The diagram of the magnetic field versus spin period. The star in the figure represents the B-P position of NuSTAR J095551+6940.8, which is plotted with our computation of $B=4.5\times10^{12}\rm\,G$, $\dot{M}=5.0\times10^{18}\rm\,g\,s^{-1}$ and the observed spin period $P=1.37\rm\,s$. The dash line represents the simulated evolutionary path of NuSTAR J095551+6940.8 with the initial field of $3.0\times10^{14}\rm\,G$ and spin period of 100\,s, as depicted in the figure. The dash-dotted line and solid line are the spin up line with the accretion rate $\dot{M}=5.0\times10^{18}\rm\,g\,s^{-1}$ and the Eddington accretion rate, respectively.
With the data from ATNF pulsar catalogue (Manchester et al. 2005), we plot the B-P position of observed binary pulsars with dots and two isolated millisecond pulsars above the Eddington spin up line with triangles.
}\label{md1850}
\end{figure}

The spin period of the observed magnetar is usually about 10 second with the estimated age $10^4\rm\,yr$ (Mereghetti 2008; Bisnovatyi-Kogan \& Ikhsanov 2014; Mereghetti et al. 2015). Since
the companion mass of NuSTAR J095551+6940.8 is more than $5.2\ms$, its main sequence stage before evolving to feed the accretion disk would last for about one million years, during which the magnetar might spin down to about 100 seconds due to the period and lifetime relation of the magnetic dipole emission as $P\sim\sqrt{t}$ (Shapiro \& Teukolsky 1983). So the initial spin period ($P_0$) of the accreting magnetar is chosen to be $100\rm\,s$. The initial magnetic field is set to be the usual values of magnetar as $3.0\times10^{14}\rm\,G$. Employing the evolution formula of B and P given in Section 2, the evolved path of B and P for NuSTAR J095551+6940.8 are calculated and plotted in Fig.\ref{md1850}.

In order to calculate the present magnetic field of NuSTAR J095551+6940.8, the spin period $P=1.37\rm\,s$ and period derivative $\dot{P}\approx-2\times10^{-10}\rm\,s\,s^{-1}$ are used in Eq.(\ref{pdot1}). We find that the simulated model curve in the B-P diagram (Fig.\ref{md1850}) requires the accretion rate to be $\dot{M}=5.0\times10^{18}\rm\,g\,s^{-1}$, and the obtained magnetic field is $B=4.5\times10^{12}\rm\,G$.
For such a result, the B-P position of NuSTAR J095551+6940.8 is included in its evolutionary path. From the B-P evolutionary history, the source might accrete $\sim0.005\ms$ within about $10^{5}\rm\,yr$, which feeds the field reduction and spins up the magnetar to the present state.
The fastness parameter is $\omega_{\rm s}=\Omega_{\rm s}/\Omega_{\rm k}\simeq0.46$, by which we get the parameter $\omega_{\rm s}/\omega_{\rm c}=0.51$ under the condition of the critical value $\omega_{\rm c}$ being $0.9$ (Ghosh \& Lamb 1992). The value of the fastness parameter is less than $\omega_{\rm c}$ which means NuSTAR J095551+6940.8 is now in the spin-up process. As the continuing decreasing of the spin period, $\omega_{\rm s}$ increases to the critical spin period $\omega_{\rm c}$ and the spin-up torque vanishes.

\section{Discussion and Conclusion}

\subsection{The evolutionary history of NuSTAR J095551+6940.8}

We deduce that NuSTAR J095551+6940.8 could be an evolving magnetar which is now spinning up.
With the initial magnetic field and spin period ($B_0\sim3.0\times10^{14}\rm\,G$ and $P_0\sim100\rm\,s$), the simulation shows that the current magnetic field of NuSTAR J095551+6940.8 has been deducted to $4.5\times10^{12}\rm\,G$ with $\dot{M}=5.0\times10^{18}\rm\,erg\,s^{-1}$ in about $10^5\rm\,yr$, while the accretion material is about $0.005\ms$.
The subsequent evolution of NuSTAR J095551+6940.8 depends on its companion mass. In the binary system, the B-P evolution of a NS is mainly affected by the companion mass, which is associated with the accretion lifetime (Pan et al. 2013).
The companion mass of NuSTAR J095551+6940.8 is more than $5.2M_{\odot}$ (Bachetti et al. 2014),
which will lead to two kinds of possible evolutionary descendants:

(a) If the companion is a $5.2-8M_{\odot}$ star, the disk
accretion can last $\sim 10^7\rm\,yr$. The accretion material will
be about $0.2M_{\odot}$ which can make NuSTAR J095551+6940.8 to be
a millisecond pulsar with the polar magnetic field of
$10^{9}\rm\,G$ and the spin period about $10\rm\,ms$, while the
super-strong out-polar field is squeezed into the NS core region.
There will be a millisecond pulsar and a heavy white dwarf left in
the binary system after the accretion.

(b) If the companion is a $>8M_{\odot}$ star, the NS may accrete about $\sim 0.02\ms$ within
$\sim10^6\rm\,yr$. A double NS system will {\bf exist} at the end of the accretion, where the recycled pulsar is similar to the Hulse-Taylor pulsar PSR 1913+16 with the dipole magnetic field of $\sim10^{10}\rm\,G$ and spin period of $\sim50\rm\,ms$.

\subsection{The B-P path of NuSTAR J095551+6940.8}

The evolutionary track of NuSTAR J095551+6940.8 in the B-P diagram above the Eddington spin-up line has been simulated (the dash line in Fig.\ref{md1850}).
From the theoretical consideration, the distribution of binary pulsars should be below the Eddington spin up line (Pan et al. 2015), since the minimum spin period of binary pulsar is arisen by the equilibrium period line with the Eddington accretion rate. However, the radio observations show that 9 binary pulsars (out of 250 ones) lie above the Eddington spin-up line (see Fig.\ref{md1850} in this paper, also see Fig.4 in Pan et al. 2013). We proposed that some of the 9 binary pulsars have experienced the similar evolutions with the super-Eddintion accretion to that of NuSTAR J095551+6940.8. That is to say, some of them may be the descendants of magnetars after accretion.
Moreover, the evolved track of NuSTAR J095551+6940.8 can also be used to explain the formation of the millisecond pulsar with the magnetic field as high as $\sim10^{9.5}\rm\,G$, e.g., two isolated millisecond pulsars B1821-24A and B1820-30A marked in Fig.\ref{md1850}, whose companions are believed to be evaporated after formations (Hessels 2008; Kluzniak 1988).

Now, we may have a whole picture of the evolutionary history of accreting NS in the B-P diagram: NS in high mass X-ray binary system and NS in low mass X-ray binary system evolve to the recycled pulsars below the Eddington spin-up line, while the descendants of ultra-luminous accreting magnetars stop at the B-P positions above the Eddington spin-up line, from where they begin their further spin-down evolutions by electromagnetic emission.

\subsection{The birth rate of ultra-luminous NS in HMXBs }

The birth rate of magnetar is about $10\%$ compared with that of normal NSs (Lorimer 2008), so the number of accreting magnetar systems shown as the ultra-luminous neutron star in HMXBs should be also $\sim10\%$ of the normal NS in HMXBs. Moreover, from the radio observations, only 9 of 250 binary pulsars lie above the Eddington spin-up line, thus we infer that the ratio of accreting magnetars to NS in X-ray binaries should be at the scale of 4\%. Until now, about a hundred of X-ray NS binary systems are observed (Liu et al. 2007). Therefore, it can be inferred that there would be a couple of ULXs to be accreting magnetars that are similar to NuSTAR J095551+6940.8.

\subsection{The possible magnetic structure of ultra luminous NS in HMXB}

Based on the model of accretion induced polar magnetic field
decay of the NS (Zhang \& Kojima 2006), the polar magnetic field
would decay one half order of magnitude after accreting $0.0001\ms$,
e.g., from $\sim10^{13}\rm\,G$ to $\sim10^{12.5}\rm\,G$. At the
same time, the out-polar field is slightly increased, which keeps
almost the same as the initial value of $\sim10^{13}\rm\,G$. For
the model acquires the magnetic flux conserving globally, the
dilution of the polar field lines will be compensated by the
expanding out-polar field lines. The strong out-polar field lines will be
totally squeezed into the NS core after accreting the mass about 0.1\ms.

The result shows that the polar field of NuSTAR J095551+6940.8 decays about two magnitude
orders, from $\sim10^{14.5}\rm\,G$ to $\sim 4.5\times10^{12}\rm\,G$ after accreting 0.005\ms. The out-polar field is almost unchanged compared with the initial value of $\sim10^{14.5}\rm\,G$. The difference of the magnetic field strength  between the polar region and equator region will
intensely arise the asymmetry of the magnetic field structure of the NS, which makes the accretion structure of the evolved magnetar in HMXB far from the spherical geometry. The Eddington luminosity, which requires a condition of the spherical accretion (Shapiro \& Teukolsky 1983; Frank et al. 2002), will be destroyed by the exotic magnetic field structure.
In addition, the super-strong multipole field structure induces the reduction of the scattering cross section of photon-electron, and in turn increases the Eddington critical luminosity (Eksi et al 2015; Laycock et al. 2015).

\subsection{Origin of the ultra luminosity of magnetar in HMXB}

In the simulation, the accretion rate $\dot{M}=5.0\times10^{18}\rm\,g\,s^{-1}$ is obtained by
consisting the observed spin period and derivative of NuSTAR J095551+6940.8 with the calculation values.
To settle the ultra luminosity difficulty, we propose two types of accretions for magnetar in HMXB, i.e., the radial random accretion $\dot{M_{\rm r}}$ without bringing orbital angular momentum and the disk accretion $\dot{M_{\rm d}}$ that contributes to the NS spin-up, thus the total accretion rate can be divided into two components,
\be \dot{M} = \dot{M_{\rm r}} + \dot{M_{\rm d}}\;.
\ee
The components of the two type accretions should be correlated to the NS magnetic structure, disk thickness and accretion environment.
For the usual NS in HMXB, the two accretion components are comparable $\dot{M_{\rm r}}\sim\dot{M_{\rm d}}$, so the source luminosity is comparable with the disk accretion luminosity $L_{\rm x}\sim L_{\rm acc}$, which is no bigger than the Eddington limit.
However, for the magnetar as an ULX in HXMB, the local super-strong field destroys the spherical accretion geometry and breaks the limit of the Eddington luminosity. %
%The random accretion dominates the emission luminosity, if  $\dot{M_{\rm r}} >> \dot{M_{\rm d}}$.
%
The structure of the super strong multipole field dominates the radiation mechanism and leads the emission to be beamed. So, the beaming factor $b$ can be introduced for the accretion luminosity $L_{\rm acc}$ that is associated with the X-ray luminosity $L_{\rm x}$: $L_{\rm acc}=bL_{\rm x}$, where $b<1$ (Feng and Soria, 2011).
For ULX NuSTAR J095551+6940.8, when assuming the pulsed fraction to be 0.05 (King and Lasota, 2016), then the accretion rate $5.0\times10^{18}\rm\,g\,s^{-1}$ of the simulation would be reasonable to infer the radial random accretion rate as $\dot{M_{\rm r}} = (1/b) \dot{M_{\rm d}} = 20 \dot{M_{\rm d}}$, which can account for the observed ultra luminosity  ($1.8\times10^{40}\rm\,erg\,s^{-1}$) of the source. Therefore, the source accretes about
$10^{-2.5}\ms$ through the polar channel as calculated from the simulation, which implies   $\sim2\times10^{-1.5}\ms$ to be accreted through the non-polar random accretion.

\section*{Acknowledgments}
This work is supported by the Strategic Priority Research Program on Space Science, the Chinese Academy of Sciences, Grant No.XDA04010300, the National Natural Science Foundation of China (NSFC 11173034), and the National Basic Research Program of China (2012CB821800).

\label{lastpage}
\end{document}